# Size effects on the second order ferroelectric phase transition in a thin film with non-symmetric boundary conditions.

K.R. Gabbasova, A.L. Pirozerskii, E.V. Charnaya

Physical Faculty, St.Petersburg State University, 198504, Ulianovskaya Street 3, Petrodvorets, St. Petersburg, Russia

### Abstract

The analytical solution is considered for the phenomenological theory of the second order ferroelectric or ferroelastic phase transition in a thin film with arbitrary boundary conditions. The general phase-plane portrait for the relevant Euler-Lagrange equation was analyzed. The order parameter distribution in the film was found for some particular sets of extrapolation lengths. The case of extrapolation lengths with opposite signs was also considered numerically. It was shown that the size effect on the order parameter and transition temperature is remarkably weakened when the extrapolation lengths have similar absolute values and opposite signs.

## 1. INTRODUCTION

Ferroelectric nanostructures attract now great attention due to the rapid increase in their practical applications [1]. Experiments on ferroelectric nanostructures showed remarkable alterations in their properties, in particular, the phase transition temperatures and spontaneous polarization compared to those in relevant bulk materials (see [2-5] and references therein). It is commonly believed that such changes are primarily induced by size reduction and increased influence of surface. Therefore, many efforts were focused on theoretical analysis of size effects on the ferroelectric and other transitions in low-dimensional systems of various geometry, especially in thin films ([6-19] and references therein). Later the theoretical models were extended over an ensemble of ferroelectric nanoparticles [20]. In [21] the effect of strains produced by substrates was emphasized.

First theoretical models for thin films based on the Landau expansion were developed assuming identical film surfaces (see [6, 12] and references therein). The sign of the extrapolation length $\delta$ which defines the boundary conditions was found to be crucial in predictions of the thin film behavior through the phase transition. Numerical calculations showed that the transition temperature moves rather strong for thin films (or small particles) and predicted a critical thickness for positive $\delta$ when ferroelectricity vanishes even at zero temperature. Later the shift of the phase transition temperature was also analyzed for asymmetric boundary conditions implying different extrapolation lengths at the film surfaces [18, 19].

Analytical consideration of the size effects on the spontaneous polarization in the film which also allowed the evaluation of the ferroelectric phase transition temperature was made till now for several particular cases of boundary conditions: for symmetric films with both positive or both negative extrapolation lengths (see [12] and references therein) and for asymmetric films with both positive extrapolation lengths [22]. The case of extrapolation lengths with opposite signs was not analyzed.

The present paper has two aims. First, we consider analytically the order parameter distribution for the ferroelectric phase transition of the second order in thin films with general boundary conditions. The phase-plane portrait will be constructed. The variety of solutions will be discussed. Second, we find numerically the order parameter distribution and shift of the phase transition temperature in thin films with asymmetric boundaries. The main attention will be focused on the case of extrapolation lengths with opposite signs. In particular, we will show that the mean order parameter and phase transition temperature can remain close to those in bulk for the extrapolation lengths with similar absolute values and opposite signs even for very thin films.

## 2. ANALITICAL SOLUTION

### 2.1. Theoretical model.

Let us consider a ferroelectric film of thickness $2L$. We assume the spontaneous polarization to be perpendicular or parallel to the film surface. We choose an orthogonal coordinate system with the origin at the film middle and the $x$ axis perpendicular to the film surface, therefore, the polarization depends on the $x$ coordinate only due to the symmetry.

In the case of the second order phase transition the Landau expansion for the free energy density $f$ can be written in the following way [6, 12]:

$$f = \frac{\alpha}{2}\eta^2 + \frac{\beta}{4}\eta^4 + \frac{\gamma}{2}\left(\frac{\partial \eta}{\partial x}\right)^2,$$ (1)

where $\eta$ is the order parameter; $\alpha = \alpha_0(T - T_0)$; $\alpha_0$, $\beta$, $\gamma$ are positive phenomenological coefficients; $T_0$ is the phase transition temperature in the relevant bulk. We imply the charge compensation on the surfaces for the polarization perpendicular to the film, then the depolarizing field is neglected. When the polarization is parallel to the film surface, the depolarizing field does not arise.

Above, the order parameter $\eta$ was assumed to be the spontaneous polarization for the ferroelectric phase transition, but the further treatment is also valid for transition into a ferroelastic phase when $\eta$ coincides with a strain tensor component or with a combination of the strain tensor components. The only requirement is that the order parameter keeps depending only on $x$.

The free energy of the whole film $\Phi$ is a sum of the volume contribution that can be found by integrating (1) over the film and of the surface ones:

$$\Phi = \int_V f dV + \int_{S_1} \frac{\gamma}{2\delta_1}\eta^2 dS + \int_{S_2} \frac{\gamma}{2\delta_2}\eta^2 dS,$$ (2)

where $\delta_1$ and $\delta_2$ are phenomenological constants (surface extrapolation lengths) for two film surfaces $S_1$ and $S_2$, respectively. Here the coefficients of $\eta^2$ are written in such a way as to ensure the generally accepted definition of the extrapolation length.

As the order parameter depends only on $x$, we obtain the following expression for the free energy of the film section with unit surface area:

$$\Phi = \int_{-L}^{L}\left(\frac{\alpha}{2}\eta^2 + \frac{\beta}{4}\eta^4 + \frac{\gamma}{2}\left(\frac{\partial \eta}{\partial x}\right)^2\right)dx + \frac{\gamma}{2\delta_1}\eta^2\bigg|_{x=-L} + \frac{\gamma}{2\delta_2}\eta^2\bigg|_{x=L}.$$ (3)

Condition for minimum of $\Phi$ implies a boundary value problem for the non-linear differential Euler-Lagrange equation:

$$\gamma\eta'' - \alpha\eta - \beta\eta^3 = 0,$$ (4)

with two non-symmetric boundary conditions

$$\left(\eta' - \frac{\eta}{\delta_1}\right)\bigg|_{x=-L} = 0, \left(\eta' + \frac{\eta}{\delta_2}\right)\bigg|_{x=L} = 0.$$ (5)

The boundary value problem (4)-(5) was posed previously in many papers (see, for instance, [6, 12] and references therein). However, it was solved only for several particular sets of extrapolation lengths. Most of solutions were made for symmetric boundaries when the order parameter distribution is symmetric with respect to the film middle. In [21] a solution for different, but positive, extrapolation lengths was written, while the possible order parameter distribution was not analyzed.

Commonly, to solve the boundary value problem for an ordinary differential equation one should find a general solution of the equation as a function of the independent variable and a set of integration constants ( two constants for a second order equation), and then find the values of the integration constants from the boundary conditions. In our case a difficulty is that the equation (4) has different types of solutions, so its general solution cannot be given by a single formula. So, we shall investigate the boundary value problem (4)-(5) according to the following plan:

  1) we shall show that the equation (4) has a simple classical mechanics analogy;
  2) we shall classify different types of the solution of (4) using the phase-plane portrait $\eta'(\eta)$ of the corresponding dynamical system;

3) using the obvious fact that boundary conditions (5) can be depicted in the phase plane by straight lines, we shall select types of trajectories admissible by geometry of the phase portrait;

4) for the selected types of the trajectories we shall obtain analytical expressions for the order parameter distribution and a transcendental equation for the corresponding integration constants.

It is easy to see that the equation (4) is equivalent from the mathematical point of view to the equation of (one-dimensional) motion of a point particle of mass $\gamma$ at a potential

$$U(\eta) = -\frac{\alpha}{2}\eta^2 - \frac{\beta}{4}\eta^4,$$

(6)

where $\eta$ corresponds to the coordinate of the particle, and $x$ – to the time. The equation (4) can be integrated once which results in the following first order differential equation:

$$\frac{\gamma}{2}(\eta')^2 + U(\eta) = E,$$

(7)

where the integration constant $E$ means the particle energy.

Now let us consider the phase-plane portrait of the corresponding dynamical system. First, note that it is symmetric with respect to the transformations $\eta \to -\eta$ and $\eta' \to -\eta'$. Next, it follows immediately from (7) that

$$\eta' = \pm\sqrt{\frac{\beta}{2\gamma}P(\eta)},$$

(8)

where $P(\eta) = \eta^4 + 2\,\mathrm{sign}(\alpha)\eta_b^2\eta^2 + \dfrac{E}{U_m}\eta_b^4$, $U_m = \dfrac{\alpha^2}{4\beta}$ and $\eta_b = \sqrt{\dfrac{|\alpha|}{\beta}}$. Note that for $\alpha < 0$ $U_m$ coincides with the maximum of $U(\eta)$ and $\eta_b$ is a magnitude of the order parameter for the relevant bulk.

Since the phase portrait geometry is qualitatively different for the cases $\alpha > 0$ and $\alpha < 0$, we consider these cases separately.

## 2.2. The case $\alpha < 0$ (Fig. 1.).

Five different types of trajectories are possible depending on roots of the polynomial $P(\eta)$ (which depends on the value of the energy $E$).

1) $E > U_m$. In this case the particle moves from $\eta = -\infty$ to $\eta = \infty$ or conversely. A typical trajectory is denoted by 1 on Fig. 1. The travel time for the whole trajectory is finite.

The polynomial $P(\eta)$ has two pairs of conjugated roots $b_1 \pm ic_1$ and $-b_1 \pm ic_1$ ($b_1 > 0$, $c_1 > 0$). It may be shown [23] that

$$\eta = b_1 + c_1 \,\mathrm{tg}\!\left(\varphi + \frac{\theta_3}{2}\right), \quad \varphi = \mathrm{am}\!\left(\frac{\beta}{2\gamma\mu}(x - x_0)\,|\,k\right),$$

(9)

where $\mathrm{am}(u\,|\,k)$ is the elliptic amplitude function [23]; the acute angles $\theta_3$ and $\dfrac{\theta_5}{2}$, elliptic modulus $k$, and coefficient $\mu$ are defined as follows:

$$\mathrm{tg}\,\theta_3 = \frac{c_1}{b_1}, \quad \mathrm{tg}^2\frac{\theta_5}{2} = \cos\theta_3, \quad k = \sin\theta_5, \mu = \frac{\sqrt{\cos\theta_5}}{c_1},$$

(10)

and $x_0$ is arbitrary.

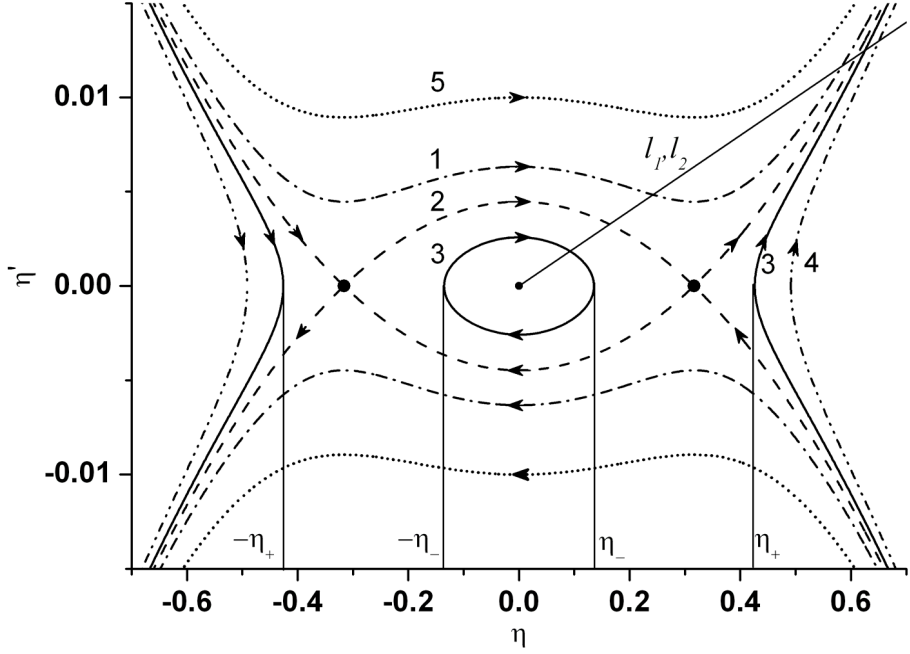

**Fig. 1.** Phase-plane portrait in the case of negative $\alpha$ ($T$=280). The phenomenological parameters of the Landau expansion are $\alpha_0$=0.01, $T_0$=300, $\beta$=2, and $\gamma$=500. The circles show the saddle points $\eta = \pm\eta_b$. The lines 1 to 5 correspond to $E$=$2U_m$, $U_m$, $U_m/3$, $-U_m$, and $5U_m$, respectively. Lines 2 are separatrices. Arrows show the direction of the motion. The straight line marked with $l_1, l_2$ is a ray corresponding to the boundary conditions (case 3 in section 2.4).

The real and imaginary parts of the roots can be expressed explicitly in terms of energy and the model parameters:

$$b_1 = \eta_b \sqrt{\frac{\sqrt{E/U_m} + 1}{2}} \,, \quad c_1 = \eta_b \sqrt{\frac{\sqrt{E/U_m} - 1}{2}} \,. \tag{11}$$

2) $E = U_m$. This case corresponds to the separatrices (dash curves 2 in Fig. 1) and two saddle points $\eta = \pm\eta_b$. Note, that while the time which is necessary for moving to (from) infinity is still finite, the one required to arrive at the saddle points is infinite. The dependence $\eta(x)$ can be explicitly expressed in hyperbolic functions.

3) $0 < E < U_m$. In this case the polynomial $P(\eta)$ has four real roots $\pm\eta_-$, $\pm\eta_+$, where

$$\eta_\pm = \eta_b \sqrt{\left|1 \pm \sqrt{1 - E/U_m}\right|} \,. \tag{12}$$

The particle either oscillates in the region $|\eta| \leq \eta_-$ or moves from $\pm\infty$ to $\pm\eta_+$ and then again moves to $\pm\infty$. A typical trajectory is denoted by 3 in Fig. 1. The dependence $\eta(x)$ can be expressed using elliptic functions.

4) $E = 0$. This case is a degenerated variant of case 3 with $\eta_- = 0$. Two possible types of trajectories are the stationary point $\eta(x) = 0$ and semi-infinite trajectories which are similar to that described above but can be expressed in trigonometric functions.

5) $E < 0$. In this case the polynomial $P(\eta)$ has two real and two imaginary roots $\pm\eta_+$ and $\pm i |\eta_-|$. Only some semi-infinite trajectories are present (curve 4 in Fig. 1). As in the case 3, $\eta(x)$ can be written in elliptic functions but the corresponding expressions are different.

### 2.3. The case $\alpha > 0$ (Fig. 2.).

In this case the potential $U(\eta)$ is a monotonic function for $\eta > 0$ and $\eta < 0$. It reaches its maximum value 0 at $\eta = 0$. Quantities $U_m$ and $\eta_b$ can not be now simply interpreted as in the case of negative $\alpha$. Note that for $E > 0$ the trajectories of the particle are infinite, but time dependency of $\eta$ is different for three energy ranges considered below.

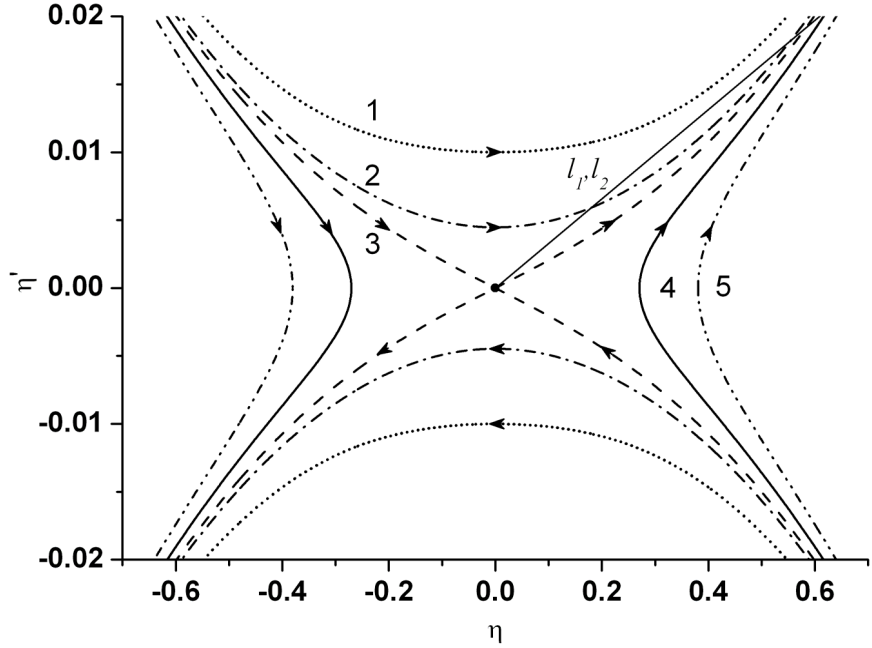

**Fig. 2.** Phase-plane portrait at in the case of positive $\alpha$ ($T = 320$). The phenomenological parameters of the Landau expansion are the same as for Fig. 1. The lines 1 to 5 correspond to $E = 5U_m$, $U_m$, 0, $-2U_m$, and $-5U_m$ respectively. Lines 3 are separatrices. Arrows show the direction of the motion. The straight line marked with $l_1 l_2$ is a ray corresponding to the boundary conditions (case 3 in section 2.4).

1) $E > U_m$ (curve 1 in Fig. 2).

As in the case 1 considered in 2.2, the polynomial $P(\eta)$ has two pairs of conjugated roots $b_1 \pm i c_1$ and $-b_1 \pm i c_1$ ($b_1 > 0, c_1 > 0$). $\eta(x)$ is given by (9), (10), but the expression (11) is now replaced by

$$b_1 = \eta_b \sqrt{\frac{\sqrt{E/U_m} - 1}{2}} \ , \ c_1 = \eta_b \sqrt{\frac{\sqrt{E/U_m} + 1}{2}} \tag{13}$$

($b_1$ and $c_1$ are exchanged).

2) $E = U_m$ (curve 2 in Fig. 2). The polynomial $P(\eta)$ has a pair of imaginary conjugated double roots $\pm i \eta_b$. The function $\eta(x)$ is defined as

$$\eta = \eta_b \ \mathrm{tg}\left(\frac{\eta_b \beta}{2\gamma}(x - x_0)\right). \tag{14}$$

3) $0 < E < U_m$. $P(\eta)$ has two pairs of imaginary conjugated roots $\pm i \eta_-$ and $\pm i \eta_+$, where $\eta_\pm$ are given by (12). For $\eta(x)$ we obtain the following expression:

$$\eta = \pm \eta_- \ \mathrm{sc}\left(\frac{\eta_+ \beta}{2\gamma}(x - x_0)\right), \tag{15}$$

where we use Glaisher's notation $\operatorname{sc} u \equiv \dfrac{\operatorname{sn} u}{\operatorname{cn} u}$ [23].

4) $E = 0$. This case corresponds to four separatrices denoted by dash curves 3 in Fig. 2 and to the saddle point $\eta = 0$.

5) $E < 0$. As in the case 5 considered in 2.2, there are two semi-infinite trajectories and $\eta(x)$ can be expressed in elliptic functions.

## 2.4. Boundary conditions.

Now let us take into account the boundary conditions (5). On the phase plane they correspond to straight lines passing through the origin. Let us denote $l_1$ and $l_2$ the lines corresponding to $x = -L$ and $x = L$, respectively. The inclination angles of $l_1$ and $l_2$ to the $\eta$ axe are $\operatorname{arctg}\left(\dfrac{1}{\delta_1}\right)$ and $-\operatorname{arctg}\left(\dfrac{1}{\delta_2}\right)$, respectively. To find a solution of the boundary value problem it is necessary first to select trajectories that intersect with both lines $l_1$ and $l_2$. Let $\eta_1$ and $\eta_2$ be corresponding to the intersection points. Then, a selected trajectory gives a solution of the boundary value problem if and only if the time required for the particle to move from $\eta_1$ to $\eta_2$ is equal to $2L$. Note that the stationary point $\eta(x) = 0$ gives the trivial solution of the boundary value problem that corresponds to the paraelectric phase and is not interesting for us. Note also that if $\eta(x)$ is a solution then $-\eta(x)$ is also a solution, and we may assume that $\eta(x) \geq 0$. (Strictly speaking, we can impose this conditions only at one point, say at $x = -L$, but the discussion of possible oscillating solutions is out of scope of the present paper, so we shall assume it for the whole film.) Taking this into account, instead of the whole lines $l_{1,2}$ now we consider only rays corresponding to $\eta \geq 0$ for which we keep the same notation.

The situation is different depending on signs of $\delta_1$ and $\delta_2$.

### 1) Positive extrapolation lengths $\delta_1 > 0$, $\delta_2 > 0$.

In this case $l_1$ passes through the first quadrant and $l_2$ – through the fourth one. Therefore, as may be seen from Figs. 1 and 2, there are no nontrivial solutions for $\alpha > 0$ while for $\alpha < 0$ only trajectories of the type with $0 < E < U_m$ are allowed. The analytical solution for this case was considered in [22] and will not be further discussed here.

### 2) Negative extrapolation lengths $\delta_1 < 0$, $\delta_2 < 0$.

In this case $l_1$ passes through the fourth quadrant and $l_2$ – through the first. For $\alpha < 0$ the admissible trajectories are selected by the following conditions:

$$E < U_m, \ \eta > \eta_+. \tag{16}$$

For $\alpha > 0$ only trajectories of the type considered in 2.3, case 5 ( $E < 0$ ) are admissible.

In all cases the condition on the time of the motion results in transcendental equations for the particle energy and second integration constant $x_0$. Detailed calculations will be given elsewhere. It should be noted that for the case of $\delta_1 = \delta_2 < 0$ the corresponding analytical solution was given earlier in [24].

### 3) Extrapolation lengths of different signs.

Due to the symmetry we may assume $\delta_1 > 0$, $\delta_2 < 0$. Then both rays $l_1$ and $l_2$ pass through the first quadrant. This case is the most complicated because, depending on the ratio of $|\delta_1|$ to $|\delta_2|$, practically all types of trajectories can be admissible for appropriate parameters of the model. In the present paper we restrict our consideration to the case $\delta_1 = -\delta_2$ when the ray $l_1$ and $l_2$ coincide with each other and an admissible trajectory must have at least two intersection points with the ray (Fig. 1). This may occur only if $\operatorname{arctg}\dfrac{\eta'}{\eta}$ is a nonmonotomic function of $\eta$. Denoting $\chi \equiv \dfrac{\eta'}{\eta}$ one obtains from (8) by a straightforward calculation:

$$\frac{d\chi}{d\eta} = \sqrt{\frac{\beta}{2\gamma}}\frac{\eta^4 - \operatorname{sign}(E)\eta_c^4}{\eta^2\sqrt{P(\eta)}},\tag{17}$$

where $\eta_c = \eta_b\sqrt[4]{\dfrac{|E|}{U_m}}$. As seen from (17), the function $\chi(\eta)$ is monotonic for $E \le 0$ and the corresponding trajectories are not admissible. Taking into account that for $\alpha < 0$ and $0 < E < U_m$ the inequality $\eta_c < \eta_b < \eta_+$ is held, we conclude that in this case only trajectories of the type considered in 2.2, case 1 are admissible. For $\alpha > 0$ all the cases 1-3 from 2.3 should be considered (Fig. 2).

Now let us consider the condition on the time of the motion. It follows from (5) and (8) that the intersection points $\eta_1$ and $\eta_2$ may be found from equations:

$$\eta_j^4 + \left(\frac{2\alpha}{\beta} - \chi_j^2\right)\eta_j^2 + \frac{4E}{\beta} = 0,\ j = 1,2,\tag{18}$$

where $\chi_j \equiv \dfrac{1}{\delta_j}$. Further computations depend on the energy magnitude.

(a) $E > U_m$.

Substituting $\eta_1$ and $\eta_2$ into (9) we obtain:

$$\eta_1 = b_1 + c_1\operatorname{tg}\left(\operatorname{am}\left(\frac{\beta}{2\gamma\mu}(-L - x_0)\,|\,k\right) + \frac{\theta_3}{2}\right),\tag{19}$$

$$\eta_2 = b_1 + c_1\operatorname{tg}\left(\operatorname{am}\left(\frac{\beta}{2\gamma\mu}(L - x_0)\,|\,k\right) + \frac{\theta_3}{2}\right).\tag{20}$$

This is a system of transcendental equations for $x_0$ and $E$. To exclude $x_0$ we rearrange (19) and (20):

$$F\left(\operatorname{arctg}\left(\frac{\eta_1 - b_1}{c_1}\right) - \frac{\theta_3}{2}, k\right) = \frac{\beta}{2\gamma\mu}(-L - x_0),\tag{21}$$

$$F\left(\operatorname{arctg}\left(\frac{\eta_2 - b_1}{c_1}\right) - \frac{\theta_3}{2}, k\right) = \frac{\beta}{2\gamma\mu}(L - x_0),\tag{22}$$

where $F(\phi, k)$ is the elliptic integral of the first kind [23]. Subtracting (21) from (22) we find an equation for E:

$$F\left(\operatorname{arctg}\left(\frac{\eta_2 - b_1}{c_1}\right) - \frac{\theta_3}{2}, k\right) - F\left(\operatorname{arctg}\left(\frac{\eta_1 - b_1}{c_1}\right) - \frac{\theta_3}{2}, k\right) = \frac{\beta L}{\gamma\mu}.\tag{23}$$

For $\alpha < 0$ two other cases should be considered.

(b) $E = U_m$.

In this case the energy is already "known". So we need only to satisfy that the following condition is fulfilled:

$$\operatorname{arctg}\left(\frac{\eta_2}{\eta_b}\right) - \operatorname{arctg}\left(\frac{\eta_1}{\eta_b}\right) = \frac{\beta\eta_b L}{\gamma}.\tag{24}$$

(c) $0 < E < U_m$.

Similarly we obtain the following equation:

$$F\left(\text{arctg}\left(\frac{\eta_2}{\eta_-}\right),k\right) - F\left(\text{arctg}\left(\frac{\eta_1}{\eta_-}\right),k\right) = \frac{\beta\eta_+ L}{\gamma}. \qquad (25)$$

At this point some comments are required. It was believed that at given parameters of the model the boundary value problem has no more than one nontrivial solution with $\eta > 0$. So, we expect that only one of the equations (23)-(25) may have a solution, and the solution, if it exists, is unique. Nevertheless, up to now no strict mathematical proof of this fact exists. Situation became even more complicated when $|\delta_1| \neq |\delta_2|$ as different types of trajectories (and even different parts of the same trajectory) may be admissible simultaneously.

## 3. NUMERICAL CALCULATIONS AND DISCUSSION

As it follows from the above consideration, the analytical solution of the boundary value problem (4)-(5) is very complicated and results at its final step in equations than can be solved only numerically. So, it is worth to find numerical solutions of this problem directly.

The order parameter variations over the film near the phase transition were found numerically using the boundary value problem solver ***bvp4c*** from MatLab 7.0. Because the solver works with systems of first order differential equations, the equation (4) was transformed in the standard way to a system of two first order ordinary differential equations. The boundary conditions (5) were also transformed to new functions. To find a numerical solution for the order parameter distribution for some sets of phenomenological parameters, film thickness, and temperature, we used a constant as an initial guess.

The following numerical parameters were used: $\alpha_0 = 0.01$, $\beta = 2$, $\gamma = 500$, and $T_0 = 300$. The relative magnitude of the equilibrium order parameter $\eta/\eta_b$ as function of $x/L$ at the temperature $T = 100$ for the film thickness $2L = 14$, fixed extrapolation length $\delta_1 = 50$, and various $\delta_2$ are shown in Fig.3.

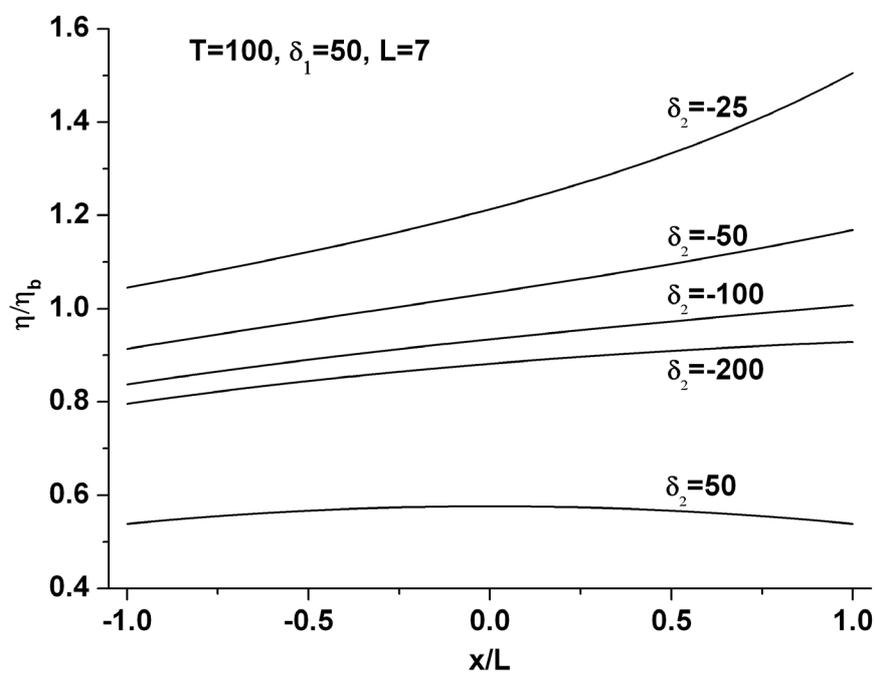

**Fig. 3.** The relative equilibrium order parameter $\eta/\eta_b$ versus $x/L$ for fixed $\delta_1 = 50$ and different $\delta_2$ shown on the panel. The film thickness $2L = 14$ and $T = 100$. Other phenomenological parameters are listed in the caption to Fig. 1.

Fig.3 emphasizes the case of different signs of the extrapolation lengths. The case $\delta_1 = \delta_2$ is shown for comparison. One can see from Fig.3 that the order parameter distribution becomes strongly asymmetric when the extrapolation lengths have different signs. For $\delta_1 = -\delta_2$, the order parameter in the film is smaller than in bulk near the boundary at $x = -L$ and increases compared to bulk near the $x = L$ boundary, the mean order parameter being close to the bulk value. The order parameter distribution remains asymmetric at different temperatures as it can be seen in Fig.4 for the case $\delta_1 = -\delta_2 = 100$.

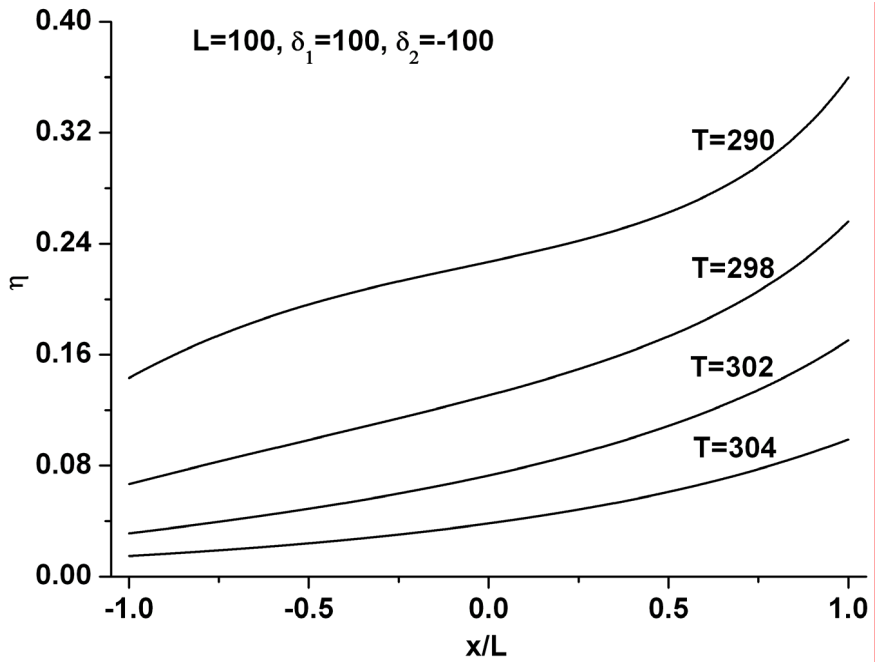

**Fig. 4.** The order parameter $\eta$ versus $x/L$ in the thin film of thickness $2L=200$ at temperatures shown on the panel. The extrapolation lengths are $\delta_1 = 100$ and $\delta_2 = -100$. Other phenomenological parameters are listed in the caption to Fig. 1.

Fig.5 shows variations of $T_C$ as a function of $L$ for three sets of extrapolation lengths.

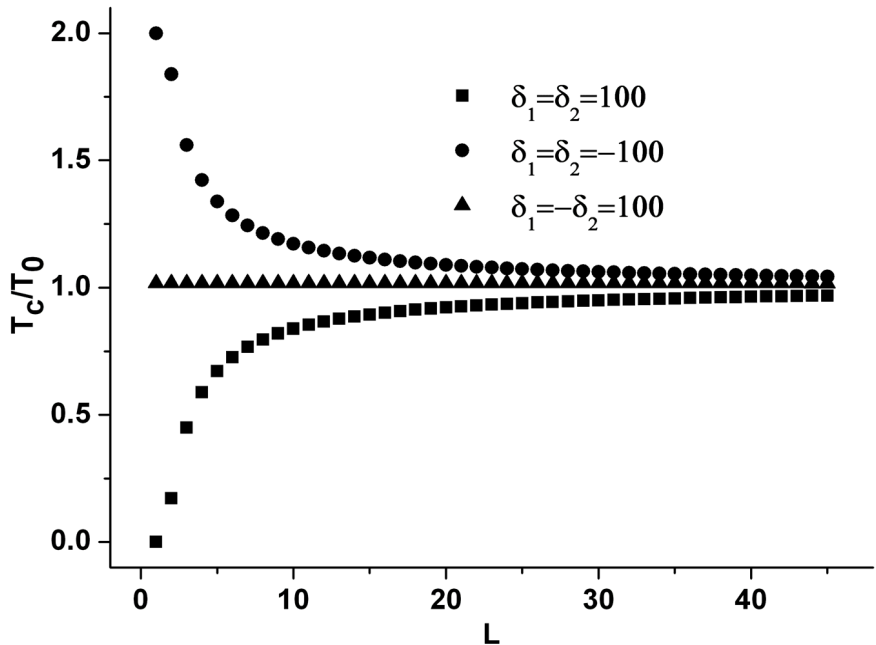

Fig. 5. The size dependence of the relative temperature $T_C/T_0$ of the phase transition for several pairs of the extrapolation lengths shown on the panel. Other phenomenological parameters are listed in the caption to Fig. 1.

This figure evidences that the competition between the extrapolation lengths with opposite signs weakens the influence of film thickness reduction on the shift of the phase transition. In fact, the size effect is hardly noticeable when the extrapolation lengths have different signs and similar absolute values. The dependences for both positive and both negative extrapolation lengths are shown for comparison.

The last result can be applied to thin ferroelectric films on substrates. When a substrate induces polarization on the adjacent film surface due to either strains produced by a lattice misfit or polarization in the substrate itself, this can be treated as the boundary condition with a negative extrapolation length. While another surface being free can correspond to a positive extrapolation length. In that case the size effect on the phase transition shift may be noticeably weakened which allows explaining the very feeble dependence of the phase transition temperature on the film thickness observed experimentally for many ferroelectric films on the substrate.

In conclusion, we considered a complete phase portrait of the Euler-Lagrange equation which corresponds to the phenomenological model of the ferroelectric or ferroelastic second order phase transition for a thin film. The analytical expressions for the order parameter distribution over the film were obtained for some sets of extrapolation lengths, in particular, for the case of extrapolation lengths with opposite signs which was not discussed previously. The numerical solutions for the order parameter and phase transition temperature were also obtained. It was found numerically that the size effect on the order parameter and transition temperature is drastically weakened when the extrapolation lengths have different signs and similar absolute values.